\documentclass{PoS}

\makeatletter
\renewcommand\speaker[1]{\if@speaker\global\@dblspeaktrue\fi
                        \global\@speakertrue
                        \global\setbox\@firstaubox
                        \hbox{{\let\thanks\@gobble
                                \let\footnote\@gobble\small 
                                \rm #1}}%
                        #1\thanks{Speakers.}\
                        }%
\makeatother

\usepackage{amssymb}
\usepackage{fontenc}
\usepackage{times} 
\usepackage{mathptmx}
\usepackage{graphicx}

\author{\speaker{Kazuharu Bamba}\\
        \llap{}KMI, Nagoya University\\}

\author{Sergei D. Odintsov$^{*\,a\,b\,c}$\footnote{Also at Tomsk State Pedagogical University, Tomsk, Russia.}\\
\llap{$^a$}Department of Physics, Nagoya University\\
Nagoya 464-8602, Japan\\
\llap{$^b$}Instituci\`{o} Catalana de Recerca i Estudis Avan\c{c}ats (ICREA),\\
Barcelona, Spain\\
\llap{$^c$}Institut de Ciencies de l'Espai (CSIC-IEEC),
Campus UAB, Facultat de Ciencies,\\
Torre C5-Par-2a pl, E-08193 Bellaterra
(Barcelona), Spain
}

\PACS{04.50.Kd, 98.80.Cq, 11.25.Mj, 04.50.Cd}

\abstract{
We review recent progress on cosmological issues and theoretical properties of 
modified gravity theories. 
In particular, we explicitly explore the conformal transformation, 
the Starobinsky inflation, and 
a unified scenario of inflation and late time acceleration 
in $F(R)$ gravity and $F(T)$ gravity (extended teleparallel gravity). 
Furthermore, we examine neutron stars and the hyperon problem 
in $F(R)$ gravity. 
Moreover, for loop quantum cosmology (LQC), the natures of finite-time future singularities in $F(T)$ gravity are presented. 
In addition, we investigate $F(T)$ gravity theories from the Kaluza-Klein (KK) and Randall-Sundrum (RS) theories. 
}

\title{Universe acceleration in modified gravities: $F(R)$ and $F(T)$ cases} 
\FullConference{KMI International Symposium 2013 on gQuest for the Origin of Particles and the Universe",\\
		11-13 December, 2013\\
		Nagoya University, Japan}
\ShortTitle{Universe acceleration in modified gravities} 

\begin{document}

%
%
\makeatletter
\setbox\@firstaubox\hbox{\small K. Bamba, S. D. Odintsov}
\makeatother

\section{Introduction}

Quite precise cosmological observations such as Type Ia Supernovae have revealed the current accelerated expansion of the universe. 
For the homogeneous and isotropic universe, 
there are two representative ways of explaining this phenomenon. 
First is the introduction of dark energy with negative pressure in general relativity. 
Second is the modification of gravity on the long distance, e.g., $F(R)$ gravity~\cite{Capozziello:2002rd, Carroll:2003wy, Nojiri:2003ft} (for reviews on dark energy and modified gravity, see, for instance,~\cite{Nojiri:2010wj, Nojiri:2006ri, Capozziello:2011et, Bamba:2012cp}). 

In this article, among various modified gravity theories, we 
concentrate on $F(R)$ gravity and $F(T)$ gravity with $T$ 
the torsion scalar in teleparallelism~\cite{A-P} 
and review the recent developments on cosmological issues and 
their theoretical properties. 
We use units of $k_\mathrm{B} = c = \hbar = 1$ and denote the
gravitational constant $8 \pi G$ by
${\kappa}^2 \equiv 8\pi/{M_{\mathrm{Pl}}}^2$
with the Planck mass of $M_{\mathrm{Pl}} = G^{-1/2} = 1.2 \times 
10^{19}$\,\,GeV. 

The article is organized as follows. 
In Sec.\ II, we review various cosmological subjects in $F(R)$ gravity. 
First, we explain the action and derive the gravitational equations. 
We explore the conformal transformation from the Jordan frame to 
the Einstein frame and its inverse transformation. 
After that, we investigate $R^2$ inflation, namely, the so-called Starobinsky inflation. Also, we present a unified model of inflation and dark energy era, and describe the properties of the finite-time future singularities. 
Moreover, we study neutron stars and the issue of hyperon. 
For the recent study of stars in dilaton gravity, see~\cite{Fiziev:2014eba}. 
Next, in Sec.\ III, we explain significant cosmological issues in $F(T)$ gravity. To begin with, we present the formulation of teleparallelism. 
We state the impossibility of the conformal transformation from $F(T)$ gravity in the Jordan frame into pure teleparallel gravity with a scalar field, 
which corresponds to the Einstein frame in the ordinary curvature gravity.  
In addition, 
we examine the analogue of the Starobinsky inflation, i.e., $T^2$ inflation. 
Furthermore, 
we illustrate the unification of inflation in the early universe 
and the late time cosmic acceleration at the dark energy dominated stage. 
We also mention the finite-time future singularities in loop quantum cosmology (LQC). 
Moreover, we study a conformally invariant $F(T)$ gravity theory. 
We further explore trace-anomaly driven inflation. 
In Sec.\ IV, we review our main results on $F(T)$ gravity theories from the Kaluza-Klein (KK) and Randall-Sundrum (RS) theories in Ref.~\cite{Bamba:2013fta}. 
In Sec.\ V, summary are described. 

\section{$F(R)$ gravity}

We present the brief review of $F(R)$ gravity. 
The action describing $F(R)$ gravity with matter is represented by 
\begin{equation}
S = \int d^4 x \sqrt{-g} \frac{F(R)}{2\kappa^2} 
+ \int d^4 x \mathcal{L}_{\mathrm{matter}} 
(g_{\mu\nu}, \Phi_{\mathrm{matter}})\,. 
\label{eq:1}
\end{equation} 
Here, $g$ is the determinant of the metric tensor $g_{\mu\nu}$, 
$\mathcal{L}_{\mathrm{matter}}$ is the Lagrangian of matter, 
and $\Phi_{\mathrm{matter}}$ denotes matter fields. 
Variation of the action in Eq.~(\ref{eq:1}) with respect to 
$g_{\mu\nu}$ leads to the gravitational equation 
\begin{equation}
F'(R) \left( R_{\mu\nu}-\frac{1}{2} g_{\mu\nu}R \right) 
= \kappa^2 T^{\mathrm{matter}}_{\mu \nu} 
-\frac{1}{2} g_{\mu \nu} \left( F'(R) R - f \right)
+ \nabla_{\mu}\nabla_{\nu} F'(R) -g_{\mu \nu} \Box F'(R)\,,
\label{eq:2}
\end{equation}
where the prime means the derivative with respect to $R$ as 
$F'(R) \equiv d F(R)/dR$, 
${\nabla}_{\mu}$ is the covariant derivative, 
$\Box \equiv g^{\mu \nu} {\nabla}_{\mu} {\nabla}_{\nu}$
is the covariant d'Alembertian, 
and $T^{\mathrm{matter}}_{\mu \nu}$ 
the energy-momentum tensor of matter. 

We suppose the flat Friedmann-Lema\^{i}tre-Robertson-Walker (FLRW) metric 
$ds^2=-dt^2+a^2(t) \sum_{i=1}^3 \left(dx^{i}\right)^2$ with $a(t)$ the scale factor. In this background, if matter is a perfect fluid, 
it follows from Eq.~(\ref{eq:2}) that the gravitational field equations read  
\begin{eqnarray} 
&&
3F'(R) H^2 
= \kappa^2 \rho_{\mathrm{matter}} +\frac{1}{2} \left( F'(R) R - F(R) \right) 
-3H\dot{F}'(R)\,,
\label{eq:3.2} \\ 
&&
-2F'(R) \dot{H}  
= \kappa^2 \left( \rho_{\mathrm{matter}} + P_{\mathrm{matter}} \right)
+\ddot{F}'(R) - H\dot{F}'(R)\,. 
\label{eq:3.3}
\end{eqnarray} 
Here, $H=\dot{a}/a$ is the Hubble parameter, and 
the dot denotes the time derivative. 
Moreover, $\rho_{\mathrm{matter}}$ and $P_{\mathrm{matter}}$ are 
the energy density and pressure of matter, respectively.

\subsection{Conformal transformation}

We make conformal transformation on the action in Eq.~(\ref{eq:1}). 
Using an auxiliary field $\chi$, 
the action in Eq.~(\ref{eq:1}) is expressed as 
\begin{equation} 
S = \int d^4 x \sqrt{-g} \frac{1}{2\kappa^2} 
\left[
F(\chi) + \left(R-\chi \right) \frac{d F(\chi)}{d \chi} \right] 
+\int d^4 x \mathcal{L}_{\mathrm{matter}} 
(g_{\mu\nu}, \Phi_{\mathrm{matter}})\,. 
\label{eq:FR9-2-IVB2-01} 
\end{equation} 
Varying this action with respect to $\chi$ leads to 
$\left(R-\chi \right) d^2 F(\chi)/d \chi^2 = 0$. 
Assuming $d^2 F(\chi)/d \chi^2 \neq 0$, we have $R = \chi$. 
By substituting this relation into the action in Eq.~(\ref{eq:FR9-2-IVB2-01}), 
we see that the original action in Eq.~(\ref{eq:1}) is recovered. 
Through the conformal transformation 
$g_{\mu\nu} \to \hat{g}_{\mu\nu} \equiv \Omega^2 g_{\mu\nu}$ with 
$\Omega^2 = \mathcal{A}$, where $\mathcal{A} \equiv d F(\chi)/d \chi$
we acquire~\cite{Maeda:1987xf, Maeda:1988ab, F-M}
\begin{eqnarray} 
&&
\hspace{-5mm}
S = \int d^4 x \sqrt{-\hat{g}} \left( \frac{\hat{R}}{2\kappa^2} 
-\frac{1}{2} \hat{g}^{\mu\nu} \partial_{\mu} \phi \partial_{\nu} \phi 
- V(\phi) \right) 
+\int d^4 x \mathcal{L}_{\mathrm{matter}} 
(\mathcal{A}^{-1} \hat{g}_{\mu\nu}, \Phi_{\mathrm{matter}})\,, 
\label{eq:ADD-FT-trace-1-a-III-B-02} \\ 
&&
\hspace{-5mm}
V(\phi) \equiv \frac{\mathcal{A} \hat{R} - F}{2 \kappa^2 \mathcal{A}^2}\,,
\quad 
\phi \equiv \sqrt{\frac{3}{2}} \frac{1}{\kappa} \ln \mathcal{A} \,.
\label{eq:ADD-FT-trace-1-a-III-B-03} 
\end{eqnarray} 
Here, the hat shows quantities in the Einstein frame. 

Next, by following the investigations in Ref.~\cite{Capozziello:2005mj}, we explore the inverse conformal transformation from the Einstein frame 
to the Jordan frame. 
Here, for simplicity, we consider the system consisting of gravity and 
a scalar field parts without other matter. 
{}From the action in Eq.~(\ref{eq:ADD-FT-trace-1-a-III-B-02}), we have 
\begin{equation}
S = \int d^{4}x \sqrt{-\hat{g}}
\left(
\frac{\hat{R}}{2\kappa^2} - \frac{1}{2} \omega \left( \psi \right)
\hat{g}^{\mu\nu} {\partial}_{\mu} \psi {\partial}_{\nu} \psi
-W(\psi) \right)\,,
\label{eq:KMI2013-1-IIC-01}
\end{equation}
with the coefficient function $\omega (\psi)$ for the kinetic term of the scalar field $\psi$ and $W(\psi)$ the potential of $\psi$. 
We introduce other scalar field $\sigma$, defined by 
$\sigma \equiv \int d\psi \sqrt{\left| \omega(\psi) \right|}$, 
and represent the action in Eq.~(\ref{eq:KMI2013-1-IIC-01}) as 
%
$S = \int d^{4}x \sqrt{-\hat{g}}
\left[
\hat{R}/\left(2\kappa^2\right) \mp \left(1/2\right) 
\hat{g}^{\mu\nu} {\partial}_{\mu} \sigma {\partial}_{\nu} \sigma 
-\tilde{W}(\sigma) \right]
$. 
%
Here, for the positive (negative) sign of $\omega (\psi)$, 
that of the kinetic term is `$-$' (`$+$'), namely, the action describes 
the non-phantom (phantom) phase, and $\tilde{W}(\sigma)$ is the potential of 
$\sigma$. 
We examine the action with the `$-$' sign kinetic term, i.e., $\sigma$ is 
a canonical scalar field. 
We make the inverse conformal transformation: 
$\hat{g}_{\mu\nu} \to \exp \left( \pm \sqrt{2/3} \kappa \sigma \right) g_{\mu\nu}$, so that the kinetic term of $\sigma$ can be removed from the action 
shown above. 
Accordingly, we obtain 
\begin{equation}
S = \int d^{4}x \sqrt{-g}
\left[ \frac{\exp \left( \pm \sqrt{2/3} \kappa \sigma \right)}{2\kappa^2} R 
- \exp \left( \pm 2\sqrt{\frac{2}{3}} \kappa \sigma \right) \tilde{W}(\sigma) \right]\,.
\label{eq:KMI2013-1-IIC-03}
\end{equation}
This is considered to be the action in the Jordan frame. 
Moreover, the equation of motion for $\sigma$ is given by 
%
$R = 2 \kappa \exp \left( \pm \sqrt{2/3} \kappa \sigma \right) 
\left( 2\kappa \tilde{W}(\sigma) \pm \sqrt{3/2} 
d \tilde{W}(\sigma)/d \sigma \right)$. 
%
This can be solved as $\sigma = \sigma (R)$. 
As a consequence, we acquire the action for an $F(R)$ gravity theory in 
Eq.~(\ref{eq:1}) without matter action, where the form of $F(R)$ 
is expressed as 
%
$F(R) = \exp \left( \pm \sqrt{2/3} \kappa \sigma (R) \right)  R 
- 2\kappa^2  \exp \left( \pm 2\sqrt{2/3} \kappa \sigma(R) \right) \tilde{W}(\sigma(R))$. 

It is important to point out that 
in the presence of matter, we have matter which couples with some function 
of $R$, and then it may be impossible to get a solution of $\phi$ as 
$\phi (R)$, as it will depend also on $\mathcal{L}_{\mathrm{matter}} 
(g_{\mu\nu}, \Phi_{\mathrm{matter}})$. 

\subsection{$R^2$ inflation (the Starobinsky inflation)}

The action for an $R^2$ theory without matter is written as 
\begin{equation}
S = \int d^4 x \sqrt{-g} \frac{1}{2\kappa^2} 
\left(R+\frac{1}{6M_{\mathrm{S}}^2} R^2 \right)\,,  
\label{eq:ADD-FT-trace-1-a-III.1}
\end{equation}
with $M_{\mathrm{S}}$ a mass scale.  
For $R^2$ inflation model~\cite{Starobinsky:1980te}, from Eq.~(\ref{eq:ADD-FT-trace-1-a-III-B-03}) we obtain~\cite{Maeda:1987xf, Maeda:1988ab, F-M}
\begin{equation}
V(\phi) = \frac{3 M_{\mathrm{s}}^2}{4 \kappa^2} 
\left[1-\exp \left(-\sqrt{\frac{2}{3}} \kappa \phi \right) \right]^2 \,.
\label{eq:ADD-FT-trace-1-a-III-B-04}
\end{equation}

The PLANCK analysis shows the spectral index $n_{\mathrm{s}} = 0.9603 \pm 0.0073$ (68\% CL) for the scalar mode of curvature perturbations and the tensor-to-scalar ratio $r < 0.11$ (95\% CL)~\cite{Ade:2013lta}. 
Provided that 
$\phi \gg \phi_{\mathrm{f}}$, where $\phi_{\mathrm{f}}$ is the value of 
$\phi$ at the end of inflation $t =t_{\mathrm{f}}$. 
When the number of \textit{e}-folds $N_{\mathrm{ar}} 
\equiv \int_{t}^{t_{\mathrm{f}}} H dt = 
\left(3/4\right) \exp \left( \sqrt{2/3} \kappa \phi \right) 
=50$ from the end of inflation to the time when the curvature perturbation with the comoving wave number $k=k_{\mathrm{cross}}$ crossed the horizon, 
for the super-horizon modes, i.e., $k \ll a H$, 
we have $n_{\mathrm{s}} 
\equiv 1 + d \ln \Delta_{\mathcal{R}}^2 (k)/d \ln k 
\big|_{k=k_{\mathrm{cross}}} 
= 1 - 6 \epsilon + 2\eta \simeq 1 - 2/N_{\mathrm{ar}} 
= 0.96$ and 
$r = 16 \epsilon = 12/N_{\mathrm{ar}}^2 =4.8 \times 10^{-3}$, 
the value of which 
is much smaller than the upper limit ($r<0.11$) suggested by PLANCK. 
Here, $\Delta_{\mathcal{R}}^2$ is the amplitude of scalar modes of the primordial curvature perturbations at $k=0.002 \, \mathrm{Mpc}^{-1}$~\cite{Hinshaw:2012aka, Salopek:1988qh}. 

It is remarkable to note that in the presence of matter, 
through the conformal transformation, 
the matter Lagrangian in Eq.~(\ref{eq:1}) would transformed into 
the form in Eq.~(\ref{eq:ADD-FT-trace-1-a-III-B-02}). 

In Ref.~\cite{Sebastiani:2013eqa}, referring to the success of 
$R^2$ inflation (the Starobinsky inflation), 
the reconstruction of an $F(R)$ gravity theory from a scalar field theory 
in the Einstein frame, which has an appropriate potential 
to realize inflation satisfying the observational constraints 
found by PLANCK. As an example, 
$V(\phi) = b_0 + b_1 \exp (\sqrt{1/3} \kappa \phi) + b_2 \exp (2\sqrt{1/3} 
\kappa \phi)$ with $b_i$ $(i= 0, 1, 2)$ constants. 
By combining this expression with the first relation in~(\ref{eq:ADD-FT-trace-1-a-III-B-03}), we find $F(R)=\left[-b_1/\left(2b_0\right)\right]R+ \left[1/\left(4b_0\right)\right]R^2 + \mathcal{C}$ with $\mathcal{C} \equiv b_1^2/\left(4b_0\right) - b_2$. 
For $R^2$ inflation, since $F(R) = R+\left[1/\left(6M_{\mathrm{S}}^2 \right) \right] R^2$, we have $-b_1/\left(2b_0\right)=1$ and $b_0=b_2$. 
As a result, if $b_1 = 3M_{\mathrm{S}}^2 /\left(4\kappa^2 \right)$, 
the potential of the corresponding scalar field theory to $R^2$ inflation 
$V(\phi)$ in Eq.~(\ref{eq:ADD-FT-trace-1-a-III-B-04}) can be obtained.

\subsection{Unified scenario of inflation and dark energy era and finite-time future singularities}

The first proposal of unification of inflation and dark energy in $F(R)$ 
gravity (\cite{Cognola:2007zu, Bamba:2010ws}) has been made in Ref.~\cite{Nojiri:2003ft}. 
Its realistic extension for exponential $F(R)$ gravity was given 
in Refs.~\cite{oai:arXiv.org:1012.2280, Bamba:2012qi}. 
There has been proposed an $F(R)$ gravity theory where inflation and late time cosmic acceleration can be realized in a unified manner~\cite{oai:arXiv.org:1012.2280, Bamba:2012qi}
\begin{equation}
F(R)=R-2\Lambda\left[1 - \exp \left(-\frac{R}{\Lambda}\right) \right]
-\Lambda_\mathrm{inf}\left\{1-\exp \left[-\left(\frac{R}{R_\mathrm{inf}}\right)^n 
\right] \right\} 
+\bar{\gamma} \left(\frac{1}{\tilde R_\mathrm{inf}^{\alpha-1}}\right)
R^\alpha\,, 
\label{eq:KMI2013-BO1-3-IIC-01}
\end{equation}
with $R_\mathrm{inf}$ and $\Lambda_\mathrm{inf}$ representative values of 
the scalar curvature and cosmological constant at the inflationary 
stage, respectively, $n ( > 1)$ a natural number, 
$\bar{\gamma} ( > 0)$ a positive dimensional constant, and 
$\alpha$ a real number.  
The last term plays an important role to realize 
the graceful exit from inflation at 
the inflation scale $\tilde R_\mathrm{inf}$. 
In fact, for $\alpha > 1$ and $n > 1$, 
there is no influence of inflation 
on the evolution of the universe at the small curvature, 
and inflation does not have any effects on the stability of 
the matter dominated stage. 

It is known that 
depending on the model, in the limit $t\to t_{\mathrm{s}}$, 
there may appear finite-time future singularities which were classified 
as follows~\cite{Nojiri:2005sx}. 

(i) Type I (``Big Rip'') singularity:\ 
When $t\to t_{\mathrm{s}}$, 
$a \to \infty$, 
$\rho_{\mathrm{eff}} \to \infty$, and 
$| P_{\mathrm{eff}} | \to \infty$. 
Here, $\rho_{\mathrm{eff}}$ and $P_{\mathrm{eff}}$ are equivalent to the total energy density and pressure of the universe, respectively. 
The case that 
$\rho_\mathrm{{eff}}$ and $P_{\mathrm{eff}}$ become finite values at $t_{\mathrm{s}}$ is included. 

(ii) Type II (``sudden'') singularity:\ 
When $t\to t_{\mathrm{s}}$,  
$a \to a_{\mathrm{s}}$, 
$\rho_{\mathrm{eff}} \to \rho_{\mathrm{s}}$, and 
$| P_{\mathrm{eff}} | \to \infty$. 

(iii) Type III singularity:\ 
When $t\to t_{\mathrm{s}}$, 
$a \to a_{\mathrm{s}}$, 
$\rho_{\mathrm{eff}} \to \infty$, and
$| P_{\mathrm{eff}} | \to \infty$. 

(iv) Type IV singularity:\ 
When $t\to t_{\mathrm{s}}$,  
$a \to a_{\mathrm{s}}$, 
$\rho_{\mathrm{eff}} \to 0$, and $| P_{\mathrm{eff}} | \to 0$. 
Only higher derivatives of $H$ becomes infinity.  
This also includes 
the case that $\rho_{\mathrm{eff}}$ and/or $| P_{\mathrm{eff}} |$ 
become finite values at $t = t_{\mathrm{s}}$. 
Here, $t_{\mathrm{s}}$, $a_{\mathrm{s}} (\neq 0)$ and $\rho_{\mathrm{s}}$ 
are constants. 

It is remarkable that adding $R^2$ term to such models with finite-time future 
singularities may lead to complete removal of 
finite-time future singularities. {}From other side, 
$R^2$ term induces early-time inflation. 
Hence, we achieve two goals: unification of inflation with dark energy and 
cancellation of finite-time future singularities. 
For the detailed study of removal of finite-time future singularities in $F(R)$ gravity via adding of $R^2$ term, which induces inflation 
within a unified manner, 
one can consult Ref.~\cite{Bamba:2008ut}.

\subsection{Neutron stars and the puzzle of hyperon}

By following the investigations in Ref.~\cite{Astashenok:2014pua}, 
we describe a model of neutron stars 
and present a solution for the issue of hyperon, i.e., the recently observed 
neutron star with its mass about $2 M_\odot$ 
cannot be realized by using an equation of state (EoS) for hyperons, 
in $F(R)$ gravity. 
In the case of the action in Eq.~(\ref{eq:1}) 
with $F(R) = R + d_1 R^2 + d_2 R^3$, where $d_1$ and $d_2$ 
are constants, for a soft equation of state (EoS) for hyperon, 
it is demonstrated that 
the maximum mass of a neutron star can explains the observations of the pulsar 
PSR J1614-2230~\cite{Demorest:2010bx}, 
and that the resultant Mass-Radius relation can be compatible with the observations. 

To begin with, we write the Tolman-Oppenheimer-Volkoff (TOV) equation 
in $F(R)$ gravity. 
With the EoS for hyperons, we examine a neutron star model in the power-law 
$F(R)$ gravity model shown above. 
We draw the diagram of a Mass-Radius relation in the $F(R)$ gravity model 
and compare it with that in general relativity. 
Furthermore, 
we explore whether the maximum mass of the neutron stars 
and the relation between Mass and Radius consistent with the observations 
can be derived in $F(R)$ gravity. 

The analyzed results are summarized as follows. 
In general relativity, for a hyperon model, the maximum mass of neutron stars 
is constrained to be smaller than around two solar mass by softening the EoS for nucleons, thanks to the hyperonization. 
On the other hand, in the power-law $F(R)$ gravity model as $F(R) = R + d_1 R^2 + d_2 R^3$, which is a simple approximation of a more complicated non-linear form of $F(R)$, with a EoS for hyperons, 
the mass of neutron stars can reach around two solar mass. 
It should be emphasized that only in the central part of neutron stars 
with its high density, 
the resultant Mass-Radius relation is different from that 
in general relativity. 
Accordingly, a EoS for neutron stars is softened effectively, so that 
the relation between Mass and Radius for 
the observed neutron stars: EXO 1745-248, 4U 1608-52, and 4U 1820-30 
can be explained. 
It may be interpreted that the following three subjects are 
resolved in the framework of modified gravity: 
the maximum mass of a neutron star, the Mass-Radius relation, and 
the puzzle of hyperons. 
A future subject would be to develop the way of solving the TOV equation non-perturbatively, and that in order to execute it, it is necessary to develop more sophisticated numerical techniques and to understand the so-called chameleon mechanism, which is a shielding effect of the deviation of 
$F(R)$ gravity from general relativity as well as quantum gravity in very high density region.

\section{$F(T)$ gravity}

We introduce orthonormal tetrad components $e_A (x^{\mu})$ $(A = 0, 1, 2, 3)$ in teleparallelism~\cite{A-P}. Here, the index $A$ is used at each point $x^{\mu}$ for a tangent space of the manifold, and hence $e_A^\mu$ is the so-called vierbein, namely, a tangent vector for the manifold. 
The metric tensor is given by  
$g_{\mu\nu}=\eta_{A B} e^A_\mu e^B_\nu$ 
$(\mu, \, \nu = 0, 1, 2, 3)$, where $\mu$ and $\nu$ are 
coordinate indices on the manifold. 
Also, the inverse of the vierbein is derived from 
the equation $e^A_\mu e_A^\nu = \delta_\mu^\nu$. 
The torsion tensor is constructed as 
$T^\rho_{\verb| |\mu\nu} \equiv 
\Gamma^{(\mathrm{W}) \rho}_{\verb|   |\nu\mu} - \Gamma^{(\mathrm{W}) \rho}_{\verb|   |\mu\nu} = e^\rho_A 
\left( \partial_\mu e^A_\nu - \partial_\nu e^A_\mu \right)$, 
where $\Gamma^{(\mathrm{W}) \rho}_{\verb|   |\nu\mu} \equiv e^\rho_A \partial_\mu e^A_\nu$ is the Weitzenb\"{o}ck connection. 
The torsion scalar is represented by 
$T \equiv S_\rho^{\verb| |\mu\nu} T^\rho_{\verb| |\mu\nu} = 
\left(1/4\right)T^{\rho\mu\nu}T_{\rho\mu\nu} + \left(1/2\right) 
T^{\rho\mu\nu}T_{\nu\mu\rho}-T_{\rho\mu}^{\verb|  |\rho}
T^{\nu\mu}_{\verb|  |\nu}$. 
Here, the superpotential is defined by 
$S_\rho^{\verb| |\mu\nu} \equiv \left(1/2\right)
\left(K^{\mu\nu}_{\verb|  |\rho}+\delta^\mu_\rho \ 
T^{\alpha \nu}_{\verb|  |\alpha}-\delta^\nu_\rho \ 
T^{\alpha \mu}_{\verb|  |\alpha}\right)$ 
with 
$K^{\mu\nu}_{\verb|  |\rho} 
\equiv -\left(1/2\right) 
\left(T^{\mu\nu}_{\verb|  |\rho} - T^{\nu \mu}_{\verb|  |\rho} - 
T_\rho^{\verb| |\mu\nu}\right)$ the contortion tensor. 
By using $T$, the action for the modified teleparallel gravity with matter 
is expressed as
\begin{equation} 
S= \int d^4x |e| \frac{F(T)}{2{\kappa}^2} 
+ \int d^4 x \mathcal{L}_{\mathrm{matter}} 
(g_{\mu\nu}, \Phi_{\mathrm{matter}})\,.  
\label{eq:KMI2013-1-IIIA-0001}
\end{equation}
where $|e| = \det \left(e^A_\mu \right)=\sqrt{-g}$. 

It is important to caution that 
there does not exist the conformal transformation~\cite{Yang:2010ji}, 
through which an $F(T)$ gravity theory reduces to a scalar field theory in pure teleparallel gravity, although in general relativity, constructed by the Levi-Chivita connection, there exists such a conformal transformation of an $F(R)$ gravity theory into a scalar field theory with the Einstein-Hilbelt term. 
Therefore, we cannot examine cosmology in the corresponding action in the Einstein frame consisting of the terms of pure teleparallel gravity and a scalar field theory, which is obtained from the action of $F(T)$ gravity with the conformal transformation~\cite{Bamba:2014zra}.

\subsection{Analogue of the Starobinsky inflation ($T^2$ inflation)}

The action of $T^2$ gravity without matter term is given by 
\begin{equation} 
S = \int d^4 x |e| \frac{1}{2{\kappa}^2} \left( T+ \frac{1}{6M_{\mathrm{S}}^2} T^2 \right) 
= \int d^4 x |e| \frac{1}{2{\kappa}^2} \left[ \left(T+ 2\nabla^{\mu} T^{\rho}_{\verb| |\mu \rho}  \right) + \frac{1}{6M_{\mathrm{S}}^2} 
T^2 \right] \,. 
\label{eq:KMI2013-1-IIID-0001} 
\end{equation}
%
Here, the second equality can be satisfied because the term 
$\nabla^{\mu} T^{\rho}_{\verb| |\mu \rho}$ is a total derivative. 
Clearly, the form of this action is different from that of 
the action in Eq.~(\ref{eq:ADD-FT-trace-1-a-III.1}), so that 
there can exist the differences of cosmology derived from these actions. 
In fact, $T^2$ inflation can be the de Sitter expansion, whereas 
$R^2$ inflation cannot it~\cite{Bamba:2014zra}. 
In the flat FLRW universe, for $T^2$ gravity, we find 
the Hubble parameter at the inflationary stage 
$H_{\mathrm{inf}} = M_{\mathrm{S}}/\sqrt{3} 
= \mathrm{constant}$, 
and thus the scale factor during inflation  
$a(t) = \bar{a} \exp \left( H_{\mathrm{inf}} t \right) 
= \bar{a} \exp \left[ \left(M_{\mathrm{S}}/\sqrt{3}\right) t \right]$, 
where $\bar{a} ( > 0 )$ is a positive constant. 
In case of $R^2$ gravity, 
if $\left| \ddot{H}/\left(M_{\mathrm{S}}^2 H \right) \right| \ll 1$ and $\left| -\dot{H}^2/\left(M_{\mathrm{S}}^2 H^2 \right) \right| \ll 1$, we obtain $H = H_{\mathrm{initial}} -\left(M_{\mathrm{S}}^2 /6 \right) \left(t-t_{\mathrm{initial}} \right)$ and $a = a_{\mathrm{initial}} \left[ H_{\mathrm{initial}} \left(t-t_{\mathrm{initial}} \right) -\left(M_{\mathrm{S}}^2 /12 \right) \left(t-t_{\mathrm{initial}} \right)^2 \right]$. Here, $t_{\mathrm{initial}}$ is the initial time of inflation, and $H_{\mathrm{initial}}$ is the value of the Hubble parameter 
and $a_{\mathrm{initial}}$ is that of the scale factor at 
$t_{\mathrm{initial}}$. Consequently, the behavior of $T^2$ inflation 
is different from that of $R^2$ inflation.

\subsection{Unification of inflation and the dark energy dominated stage}

A unified model between inflation and the dark energy dominated stage 
has been proposed in Ref.~\cite{deHaro:2012zt}. 
In this subsection, we adopt the unit where $\kappa^2 = 1$. 
Provided that the universe is filled by a barotropic fluid with its equation of state $w_{\mathrm{fl}} \equiv P_{\mathrm{fl}} /\rho_{\mathrm{fl}}  = -1-f_{\mathrm{fl}} (\rho_{\mathrm{fl}})/\rho_{\mathrm{fl}}$, where 
$\rho_{\mathrm{fl}}$ and $P_{\mathrm{fl}}$ are the energy density and pressure of a barotropic fluid, respectively, and 
$f_{\mathrm{fl}} (\rho_{\mathrm{fl}})$ is a function of $\rho_{\mathrm{fl}}$. We explore $f_{\mathrm{fl}}$ bringing with two zeros, i.e., two de Sitter solutions, leading to a cosmological constant with its large value 
during inflation in the early universe and a fluid with zero pressure such as dust at the late time. 

The Lagrangian is described by $\mathcal{L} = V_\mathrm{s} F(T) -V_\mathrm{s} 
\rho_{\mathrm{fl}} (V_{\mathrm{s}})$ with $V_\mathrm{s}$ the spatial volume. 
Since the conjugate momentum is 
$\partial \mathcal{L} / \partial \dot{V}_\mathrm{s}$, 
we see that the Hamiltonian becomes 
$\mathcal{H} = \dot{V}_\mathrm{s} \left(\partial \mathcal{L} / \partial \dot{V}_\mathrm{s}\right) - \mathcal{L} = \left(2TF'(T) -F(T) + \rho_{\mathrm{fl}} \right) V_{\mathrm{s}}$, where the prime depicts the derivative with respect to 
$T$. {}From the Hamiltonian constraint in $F(T)$ gravity is the same as one in general relativity, namely, $\mathcal{H} = 0$, with $V_{\mathrm{s}} \neq 0$, 
we have $\rho_{\mathrm{fl}} = F(T) - 2F'(T) T$. By using this relation, 
we observe a curve in the plane of $(H, \rho_{\mathrm{fl}})$. 
For the pure teleparallel gravity, i.e., $F(T) = T$, with $T = -6H^2$, 
the above relation reads $T/2 + \rho_{\mathrm{fl}} = 0$, from which 
we find the Friedmann equation $H^2= \rho_{\mathrm{fl}}/3$. 
Hence, the dynamics is determined by 
the equation system $\dot{H} = f_{\mathrm{fl}} (\rho_{\mathrm{fl}})/2$ 
and $\dot{\rho}_{\mathrm{fl}} = 3 H f_{\mathrm{fl}} (\rho_{\mathrm{fl}})$. 
As an example, we build a model with a small cosmological constant 
$\lambda^4$ so that the energy density of the universe can be $\rho_{\mathrm{fl}} + \lambda^4$. In this case, if $\rho_{\mathrm{fl}} = \rho_{\mathrm{fl}}^{(\mathrm{inf})}$ is large, the fluid behaves as a large cosmological constant, whereas, when $\rho_{\mathrm{fl}}$ is small, its pressure is almost zero. 
Thus, for $P_{\mathrm{fl}} = - \rho_{\mathrm{fl}}^2/ \rho_{\mathrm{fl}}^{(\mathrm{inf})}$, since $w_{\mathrm{fl}} = - \rho_{\mathrm{fl}}/ \rho_{\mathrm{fl}}^{(\mathrm{inf})}$, we acquire $f_{\mathrm{fl}} (\rho_{\mathrm{fl}}) = - \rho_{\mathrm{fl}} \left(1 - \rho_{\mathrm{fl}}/\rho_{\mathrm{fl}}^{(\mathrm{inf})} \right)$. In this model, there exist two critical points. The first point $P_1$ is 
$(H, \rho_{\mathrm{fl}}) = (\lambda^2 /\sqrt{3}, 0)$, where 
$w_{\mathrm{fl}} = -1$ (i.e., the de Sitter expansion)
and the energy fraction of a fluid $\Omega_{\mathrm{fl}} \equiv \rho_{\mathrm{fl}} / \left(3H^2 \right) \simeq 1$. 
The second point $P_2$ is 
$(H, \rho_{\mathrm{fl}}) = (\sqrt{\left(\rho_{\mathrm{fl}}^{(\mathrm{inf})} + \lambda^4 \right)/3}, \rho_{\mathrm{fl}}^{(\mathrm{inf})})$, where 
$w_{\mathrm{fl}} = -1$ (i.e., the de Sitter expansion)
and $\Omega_{\mathrm{fl}} = 0$. 
Moreover, for $\lambda^4  \ll \rho_{\mathrm{fl}} \ll \rho_{\mathrm{fl}}^{(\mathrm{inf})}$, we see that $w_{\mathrm{fl}} \simeq 0$ and $\Omega_{\mathrm{fl}} \simeq 1$ (i.e., the matter (fluid) dominated stage). 
Therefore, it can be considered that in the present model, 
the point $P_1$ corresponds to the inflationary stage, 
the point $P_2$ describes the dark energy dominated stage, 
and there exists the matter dominated stage between the points $P_1$ 
and $P_2$. 
In comparison with the unified model in 
Eq.~(\ref{eq:KMI2013-BO1-3-IIC-01}) 
of inflation and the late time 
cosmic acceleration in $F(R)$ gravity, 
the results stated in the above $F(T)$ gravity model are reasonable, 
because also in the $F(R)$ gravity model 
in Eq.~(\ref{eq:KMI2013-BO1-3-IIC-01}), 
inflation, the stable matter dominated stage, and 
the late-time cosmic acceleration can be realized. 

In Ref.~\cite{Bamba:2012ka}, finite-time future singularities\footnote{In Ref.~\cite{Bamba:2012vg}, the finite-time future singularities in $F(T)$ gravity have been examined in detail.} have been studied in the framework of loop quantum cosmology (LQC) (for recent reviews on LQC, see~\cite{Ashtekar:2011ni, Bojowald:2012xy}). 
It has been demonstrated that in LQG, holonomy corrections add $\rho_{\mathrm{fl}}^2$ correction term to the Friedmann equation, so that a Big Rip singularity can be cured. 
Furthermore, it has been investigated whether other types of finite-time future singularities can be removed in LQC. 
As a consequence, Big Rip singularities cannot appear because 
along an ellipse in the plane of $(H, \rho_{\mathrm{fl}})$, 
the Friedmann equation moves in the anti-clockwise manner. 
However, when $f_{\mathrm{fl}}$ diverges at some energy 
density smaller than the critical one, sudden singularities can occur

\subsection{Conformally invariant $F(T)$ gravity theory}

Recently, a conformally invariant scalar field theory 
has been investigated in Ref.~\cite{Bamba:2013jqa}. 
In ordinary curvature gravity, the action of the conformally invariant scalar field theory is represented as 
\begin{equation} 
S = \int d^4x \sqrt{-g}\left(
\frac{B_1}{2} \varphi^2 R -\frac{1}{2}\nabla_{\mu} \varphi 
\nabla^{\mu} \varphi 
-\frac{\varphi^{m_1+1}}{m_1 +1}\right) \,, 
\label{eq:KMI2013-1-IIIE-00001} 
\end{equation}
with $B_1$ and $m_1$ constants. The variation of this action with respect to 
the scalar field $\varphi$ leads to the equation of motion for $\varphi$ as 
$\Box \varphi - B_1 \varphi R + \varphi^{m_1} = 0$. 
It is known in Ref.~\cite{Buchbinder:1992rb} that 
this equation is invariant under 
the conformal transformation for $\varphi$ of 
$\varphi \equiv \exp \left(\gamma_1 \eta \right) \hat{\varphi}$, 
where $\eta=\eta(\mbox{\boldmath $x$})$ and $\gamma_1$ is a constant, 
and that for $g_{\mu\nu}$ of $\hat{g}_{\mu\nu}=\exp\left({\eta(\mbox{\boldmath $x$})}g_{\mu\nu}\right)$, 
when 
$B_1 = 1/6$, $m_1 = 3$, and $\gamma_1 = -1/2$. 
By analogy with the above fact, it is considered that 
also in teleparallel gravity, 
the equation of motion for $\varphi$ is invariant under 
the conformal transformation for $\varphi$ of 
$\varphi \equiv \exp \left(\gamma_2 \eta \right) \hat{\varphi}$ 
with $\gamma_1$ a constant 
and that for $g_{\mu\nu}$ of $\hat{g}_{\mu\nu}=\exp\left({\eta(\mbox{\boldmath $x$})}g_{\mu\nu}\right)$. 
As a result, if the action is given by 
\begin{equation} 
S=\int d^4x\ e \left( -\frac{B_2}{2}\varphi^2 T 
+\frac{1}{2}\nabla_{\mu}\varphi \nabla^{\mu}\varphi 
+B_3 \varphi T^{\rho}_{\;\;\;\mu\rho}\nabla^{\mu}\varphi-\frac{\varphi^{m_2 + 1}}{m_2 + 1} \right) \,, 
\label{eq:KMI2013-1-IIIE-00004} 
\end{equation}
{}from this action, the equation of motion for $\varphi$ reads 
$\Box\varphi + B_2 \varphi T 
+ B_3 \varphi \nabla^{\mu}T^{\rho}_{\;\;\;\mu\rho} 
+ \varphi^{m_2} = 0$, 
where $B_2 = 1/6$, $B_3 = 1/3$, and $m_2 = 3$. 
Here, in deriving Eq.~(\ref{eq:KMI2013-1-IIIE-00004}), 
we have used the relation 
$R=-T-2 \nabla^{\mu}T^{\nu}_{\;\;\mu\nu}$. 
It is significant to obtain the above consequence, 
namely, the torsion scalar can non-minimally couple to a scalar field 
conformally, 
that the second term on the right-hand side of this relation is the total derivative, 
and that this total derivative term consists of both 
the torsion scalar and the Ricci scalar. 

\subsection{Trace-anomaly driven inflation}

We study trace-anomaly driven inflation by following the observations 
in Ref.~\cite{Bamba:2014zra}. 
The trace anomaly is given by~\cite{Duff:1993wm} 
$T_\mathrm{anomaly}= \tilde{c}_1 \left[\mathcal{F} + \left(2/3\right) \Box R\right] + \tilde{c}_2 \mathcal{G} + \tilde{c}_3 \Box R$. 
Here, 
$\mathcal{F} = \left(1/3\right) R^2 -2 R_{\mu\nu}R^{\mu\nu}+
R_{\mu\nu\rho\sigma}R^{\mu\nu\rho\sigma}$ 
is the square of the four-dimensional Weyl tensor, and 
$\mathcal{G}=R^2 -4 R_{\mu\nu}R^{\mu\nu}+
R_{\mu\nu\rho\sigma}R^{\mu\nu\rho\sigma}$ is the Gauss-Bonnet invariant. 
For $N$ scalars, $N_{1/2}$ spinors, $N_1$ vector fields, $N_2$ 
($=0$ or $1$) gravitons and $N_\mathrm{HD}$ higher-derivative conformal 
scalars, $\tilde{c}_1$ and $\tilde{c}_2$ are represented as 
\begin{eqnarray} 
&&
\tilde{c}_1 = \frac{N +6N_{1/2}+12N_1 + 611 N_2 - 8N_\mathrm{HD}}{120(4\pi)^2}
\,, 
\label{eq:00000a} \\
&&
\tilde{c}_2 =- \frac{N+11N_{1/2}+62N_1 + 1411 N_2 -28 N_\mathrm{HD}}{360(4\pi)^2}\,.
\label{eq:00000b}
\end{eqnarray}
For an ordinary matter (except conformal scalars with higher-derivative 
terms), $\tilde{c}_1 >0$ and $\tilde{c}_2 <0$. 
Moreover, $\tilde{c}_3$ can be taken as an arbitrary value, because 
the finite renormalization of the local counterterm $R^2$ 
can shift the value of $\tilde{c}_3$. 
In the flat FLRW universe, we have 
$R = 12 H^2 + 6 \dot H$, $\mathcal{F}=0$, 
$\mathcal{G}=24\left(H^4 + H^2 \dot H\right)$, and therefore the trace of energy-momentum tensor of $T$ reads 
$T_\mathrm{T} 
= - \left(2/\kappa^2 \right) 
\left( 6H^2 + 3\dot H - F - 3\dot H F' - 12 H^2 F' 
+ 36 H^2 \dot H F'' \right)$ with the prime meaning the derivative 
with respect to $T$. 
It follows the gravitational field equations with 
the trace anomaly that we find  
\begin{eqnarray} 
&&
0 = \frac{2}{\kappa^2} \left( F + 3\dot H F' + 12 H^2 F' 
- 36 H^2 \dot H F'' \right)
\nonumber \\
&&
{}
- \left( \frac{2}{3} \tilde{c}_1 + \tilde{c}_3 \right) \left( \frac{d^2}{dt^2} + 3 H \frac{d}{dt} \right)
\left( 12 H^2 + 6 \dot H \right) + 24 \tilde{c}_2 \left( H^4 + H^2 \dot H \right)\,.
\label{eq:add-IIID-01}
\end{eqnarray}
For the de Sitter space, we have a constant Hubble parameter 
$H=H_\mathrm{const}$, so that Eq.~(\ref{eq:add-IIID-01}) can be 
written to 
%
$0 = \left(2/\kappa^2 \right) \left( F + 12 H_\mathrm{const}^2 F' \right)
+ 24 \tilde{c}_2 H_\mathrm{const}^4$. 
%
If $\tilde{c}_2 = 0$, namely, there is no contribution of the trace anomaly, 
we obtain 
%
$F_\mathrm{EH} = - 2 T + 12 H_\mathrm{const}^2$, 
which is equivalent to general relativity with a cosmological constant. 
With $T=- 6 H_\mathrm{const}^2$, this expression 
can meet the above equation. 
For example, if $F(T) = T + \beta T^n$ with $\beta$ and $n$ are constants, 
the above equation yields 
$H_\mathrm{const}^2 = 0$ and/or 
$1+ \left(2n-1\right) \beta \left(-6H_\mathrm{const}^2 \right)^{n-1} 
+2H_\mathrm{inf}^2 \tilde{c}_2 \kappa^2 =0$. 
In case of $n=2$, provided that  
$\left(9 \beta/\kappa^2\right) - \tilde{c}_2 > 0$, 
we get a de Sitter solution. Thus, exponential de Sitter inflation 
can occur. 
Since $\tilde{c}_2 <0$ for an ordinary matter, 
if $\beta > 0$, the above relation satisfies. 
We remark that the de Sitter solution in $T^2$ gravity can be 
unstable, and thus $T^2$ inflation can end~\cite{Bamba:2014zra}.  
We also note that compared with $T^2$ gravity, 
for the action in Eq.~(\ref{eq:1}) without matter, where 
$F(R) = R + \upsilon R^{q}$ with $\upsilon$ and $q$ constants, 
the parameter region to produce the de Sitter solution is smaller. 

\section{$F(T)$ gravity theories from the Kaluza-Klein (KK) and Randall-Sundrum (RS) theories}

In Ref.~\cite{Bamba:2013fta}, four-dimensional effective $F(T)$ gravity theories is constructed from the five-dimensional Kaluza-Klein (KK)~\cite{F-M, MKK-A-C-F, Overduin:1998pn} and Randall-Sundrum (RS)~\cite{Randall:1999ee, Randall:1999vf} theories. 

\subsection{From the KK theory}

Provided that the ordinary KK reduction procedure~\cite{F-M, MKK-A-C-F, Overduin:1998pn} from the five-dimensional space-time to the four-dimensional one can also be used in $F(T)$ gravity. 
One of the dimensions of space is compactified to a small circle and 
the four-dimensional space-time is extended infinitely. 
The radius of the fifth dimension is taken to be of order of the Planck 
length in order for the KK effects not to be seen. Thus, the size of 
the circle is so small that phenomena in sufficiently low energies cannot be 
detected. 
In the following, we only explore the gravity part of the action and remove its matter part. 
The five-dimensional action for $F(T)$ gravity~\cite{Capozziello:2012zj} 
is given by
${}^{(5)}S= 
\int d^5 x \left|{}^{(5)}e\right| 
F({}^{(5)}T)/ \left(2 \kappa_5^2 \right)$ 
with the torsion scalar 
${}^{(5)}T \equiv 
\left(1/4\right) T^{a b c}T_{a b c} + \left(1/2\right)
T^{a b c}T_{c b a}-T_{a b}^{\verb|  |a}
T^{c b}_{\verb|  |c}$, 
the form of which is the same as that of the four-dimensional 
one, and the Latin indices $(a, b, \dots = 0, 1, 2, 3, 5)$ 
with ``$5$'' the fifth-coordinate component. 
Here, the superscripts of ``$(5)$'' or the subscripts of ``$5$'' 
expresses the five-dimensional quantities, 
${}^{(5)}e = \sqrt{{}^{(5)}g}$, where ${}^{(5)}g$ the determinant of 
${}^{(5)}g_{\mu\nu}$, 
and $\kappa_{5}^2 \equiv 8 \pi G_5 = 
\left( {}^{(5)}M_{\mathrm{Pl}} \right)^{-3}$, 
where $G_5$ is the gravitational constant and 
$M_{\mathrm{Pl}}^{(5)}$ is the Planck mass. 
The five-dimensional metric is represented as 
${}^{(5)}g_{\mu\nu} = \mathrm{diag} (g_{\mu\nu}, -\psi^2)$. 
Here, 
$\psi \equiv \tau/\tau_{*}$ is a dimensionless quantity and 
a homogeneous scalar field only with its time dependence. 
Also, $\tau$ is a homogeneous scalar field with a mass dimension and 
$\tau_{*}$ is a fiducial value of $\tau$. 
With $e^A_a = \mathrm{diag} (1, 1, 1, 1, \psi)$ and 
$\eta_{a b} = \mathrm{diag} (1, -1, -1, -1, -1)$,  
the four-dimensional effective action is written as 
$S_{\mathrm{KK}}^{(\mathrm{eff})} = \int d^4x |e| 
\left[1/\left( 2\kappa^2 \right) \right] \psi 
F(T +\psi^{-2} \partial_{\mu} \psi \partial^{\mu} \psi)$.

For $F(T) = T-2\Lambda_4$, where $\Lambda_4 (>0)$ is the four-dimensional positive cosmological constant, 
with a scalar field $\sigma$ as 
$\psi \equiv \left(1/4\right) \xi^2$, 
the above action is rewritten to~\cite{F-M}
$S_{\mathrm{KK}}^{(\mathrm{eff})} |_{F(T)=T-2\Lambda_4} = \int d^4x |e| 
\left(1/\kappa^2 \right) 
\left[ \left(1/8\right) \xi^2 T + \left(1/2\right) \partial_{\mu} \xi \partial^{\mu} \xi - \Lambda_4 \right]$. 
In the flat FLRW space-time with the metric 
$ds^2 = dt^2 - a^2(t) \sum_{i=1,2,3}\left(dx^i\right)^2$, 
we find $g_{\mu \nu}= \mathrm{diag} (1, -a^2, -a^2, -a^2)$ and 
$e^A_\mu = \mathrm{diag} (1,a,a,a)$. 
By using these expressions, we acquire the relation $T=-6H^2$. 
Accordingly, with the above relation, 
the gravitational field equations is derived as 
$\left(1/2\right) \dot{\xi}^2 
-\left(3/4\right) H^2 \xi^2 + \Lambda_4 = 0$ and 
$\dot{\xi}^2 + H \xi \dot{\xi} 
+\left(1/2\right) \dot{H} \xi^2 = 0$~\cite{Geng:2011aj}, 
and the equation of motion for $\xi$ reads 
$\ddot{\xi} + 3H\dot{\xi} + \left(3/2\right) H^2 \xi = 0$.  
It follows from the above gravitational field equations that 
$\left(3/2\right) H^2 \xi^2 -2\Lambda_4 + H \xi \dot{\xi} 
+ \left(1/2\right) \dot{H} \xi^2 = 0$. 
We get a solution of the Hubble parameter during inflation 
$H = H_{\mathrm{inf}} = \mathrm{constant} (>0)$ 
and $\xi = \xi_1 \left(t/\bar{t}\right) + \xi_2$, where 
$\xi_1$ and $\xi_2 (>0)$ are constants and 
$\bar{t}$ shows a time. 
In the limit of $t \to 0$, we see that 
exponential de Sitter inflation can happen as 
$H_{\mathrm{inf}} \approx \left(2/\xi_2\right) \sqrt{\Lambda_4/3}$, 
$a \approx \bar{a} \exp \left( H_{\mathrm{inf}} t \right)$, 
and $\xi \approx \xi_2$. 
Also, from the equation of motion for $\xi$, 
we obtain $\xi_1 \approx -\left(1/2\right) \xi_2 H_{\mathrm{inf}} \bar{t} 
\approx -\sqrt{\Lambda_4/3} \bar{t}$. 

\subsection{From the RS type-II model}

The RS type-I model~\cite{Randall:1999ee} consists of two branes. 
One is a brane with positive tension at $y=0$ and 
the other is that with negative tension at $y=u$. 
Here, $y$ denotes the fifth dimension. 
By using the warp factor $\exp \left( -2|y|/l \right)$ and 
the negative cosmological constant $\Lambda_5 (< 0)$ in the bulk, 
the five-dimensional metric is written as 
$ds^2 = \exp \left(-2|y|/l\right) g_{\mu\nu} (x) 
dx^{\mu} dx^{\nu} + dy^2$ with $l=\sqrt{-6/\Lambda_5}$. 
In the limit $u \to \infty$, 
we have the RS type-II model~\cite{Randall:1999vf}, namely, 
the single brane with the positive tension exists 
in the anti-de Sitter (AdS) bulk space. 
The gravitational field equation on the brane in the RS type-II model 
has been derived in Ref.~\cite{Shiromizu:1999wj}. 
Recently, this procedure with 
(a) the induced (Gauss-Codazzi) equations on the brane, 
(b) the Israel's junction conditions on it, and 
(c) $Z_2$ symmetry of $y \leftrightarrow -y$, 
has been applied to teleparallel gravity~\cite{Nozari:2012qi}. 
In comparison with the case of ordinary curvature gravity, 
there appear additional terms originating from the projection on the brane 
of the vector portion of the torsion tensor in the bulk. 

For the flat FLRW space-time, 
on the brane, the Friedmann equation reads 
$H^2 \left(d F(T)/dT \right) = -\left(1/12\right) \left[ F(T) - 4 \Lambda 
-2 \kappa^2 \rho_{\mathrm{matter}} 
- \left(\kappa_5^2/2\right)^2 \mathcal{I} \rho_{\mathrm{matter}}^2 
\right]$, where 
$\mathcal{I} \equiv \left(11-60 w_{\mathrm{matter}} +93 w_{\mathrm{matter}}^2 \right)/4$ and a EoS of matter 
$w_{\mathrm{matter}} \equiv P_{\mathrm{matter}}/\rho_{\mathrm{matter}}$. 
The contributions of teleparallel gravity, which do not exist for 
curvature gravity, are involved in $\mathcal{I}$. 
Moreover, on the brane, 
the effective cosmological constant becomes 
$\Lambda \equiv \Lambda_5 + \left(\kappa_5^2/2\right)^2 \lambda^2$, 
where $\lambda (> 0)$ is the brane tension 
and $G = \left[1/\left(3\pi \right) \right] \left(\kappa_5^2/2\right)^2 
\lambda$. 
In what follows, we examine the dark energy dominated stage, so that 
the contributions from matter, e.g., $\rho_{\mathrm{matter}}$ or 
$w_{\mathrm{matter}}$, can be neglected. 
For $F(T) = T - 2 \Lambda_5$, 
by using the relation $T=-6H^2$, 
we acquire a de Sitter solution 
$H = H_{\mathrm{DE}} = 
\sqrt{\Lambda_5+\kappa_5^4\lambda^2/6}
= \mathrm{constant}$ 
and $a(t) = a_{\mathrm{DE}} \exp \left( H_{\mathrm{DE}} t \right)$, 
where $a_{\mathrm{DE}} (>0)$ is a positive constant. 
Hence, the late time accelerated expansion of the universe can occur. 
For $F(T) = T^2/\bar{M}^2 + \zeta \Lambda_5$, 
where $\bar{M}$ a mass scale and $\zeta$ is a constant, 
we get $H = H_{\mathrm{DE}} = 
\left\{ \left(\bar{M}^2/108\right) 
\left[ \left(\zeta-4\right) \Lambda_5 
-4 \left(\kappa_5^2/2\right)^2 \lambda^2 
\right] \right\}^{1/4} 
= \mathrm{constant}$. 
Here, since the content of the 4th root should be positive or 
zero, we see that 
$\zeta \geq 4 + \left(\kappa_5^2 \lambda^2\right)/\Lambda_5$. 

\section{Summary}

In the present article, 
we have reviewed recent progress on issues of cosmic acceleration 
and theoretical natures of $F(R)$ gravity and $F(T)$ gravity. 
In the former part, 
we have described various cosmological and theoretical problems 
in $F(R)$ gravity. 
We have examined the conformal transformation from $F(R)$ gravity 
(i.e., the Jordan frame) to general relativity with 
a scalar field (i.e., the Einstein frame) and its inverse version. 
In addition, 
we have studied $R^2$ inflation (the Starobinsky inflation). 
Also, we have explained a unified scenario of inflation and the dark energy 
dominated stage. 
Moreover, we have presented the classification of the finite-time future singularities. 
Furthermore, we have examined neutron stars and the hyperon issue. 
Next, in the latter part, 
we have stated a number of cosmological issues as well as theoretical properties in $F(T)$ gravity. First of all, we have written the basic formulation of teleparallelism. 
Then, we have remarked the impossibility of the conformal transformation from the Jordan frame to the Einstein frame. 
We have further investigated the analogue of the Starobinsky inflation, namely, $T^2$ inflation. 
In addition, 
we have demonstrated the unification between inflation  
and the late time accelerated expansion of the universe. 
Furthermore, we have mentioned the finite-time future singularities in loop quantum cosmology (LQC). 
Moreover, 
we have shown a conformally invariant $F(T)$ gravity theory. 
Finally, 
we have reviewed our main results of Ref.~\cite{Bamba:2013fta} 
in terms of derivations of $F(T)$ gravity theories from the Kaluza-Klein (KK) and Randall-Sundrum (RS) theories. 

It could be expected that through the investigations on various theoretical aspects of a number of gravity theories, we can acquire some clues to resolve the issues of cosmic acceleration including the mechanism of inflation in the early universe and the origin of dark energy at the present time.

\section*{Acknowledgments}

The work is supported in part
by the JSPS Grant-in-Aid for
Young Scientists (B) \# 25800136 (K.B.);\
that for Scientific Research
(S) \# 22224003 and (C) \# 23540296 (S.N.);\ 
and Basic Sciences Task of Ministry of Education and Science (Russia) 
(S.D.O.). 


\end{document}